\documentclass[12pt]{iopart}

\usepackage{iopams}  
\usepackage{graphicx}

\newcommand{\text}[1]{\mathrm{#1}} 
\newcommand{\eqref}[1]{(\ref{#1})}
\newcommand{\Eqref}[1]{Eq.~\eqref{#1}}

\newcommand{\fss}[1]{#1\!\!\!/}   
\newcommand{\fsl}[1]{#1\!\!\!\!/} 

\begin{document}

\title{External Fields as a Probe for Fundamental Physics}

\author{Holger Gies}

\address{Institute for Theoretical Physics, Heidelberg
  University,\\ Philosophenweg 16, D-69120 Heidelberg, Germany }
\ead{h.gies@thphys.uni-heidelberg.de}
\begin{abstract}
  Quantum vacuum experiments are becoming a flexible tool for investigating
  fundamental physics. They are particularly powerful for searching for new
  light but weakly interacting degrees of freedom and are thus complementary
  to accelerator-driven experiments. I review recent developments in this
  field, focusing on optical experiments in strong electromagnetic fields. In
  order to characterize potential optical signatures, I discuss various low-energy
  effective actions which parameterize the interaction of particle-physics
  candidates with optical photons and external electromagnetic
  fields. Experiments with an electromagnetized quantum vacuum and optical
  probes do not only have the potential to collect evidence for new physics,
  but special-purpose setups can also distinguish between different
  particle-physics scenarios and extract information about underlying
  microscopic properties. 
\end{abstract}


\section{Introduction}

With the advent of quantum field theory, our understanding of the
vacuum has changed considerably from a literal ``nothing'' to such a
complex ``something'' that its quantitative description requires to
know almost ``everything'' about a given system.

Consider a closed quantum field theoretic system in a box with boundaries,
where all matter density is already removed (pneumatic vacuum). Still, the
walls of the system which are in contact with surrounding systems may have a
temperature, releasing black-body radiation into the box. Charges and
currents outside the box can create fields, exerting their influence on the
box's inside. The box may furthermore be placed on a gravitationally curved
manifold. And finally, the boundaries itself do generally impose constraints
on the fluctuating quantum fields inside.

A pure quantum vacuum which is as close to trivial as possible
requires to take the limit of vanishing parameters which quantify the
influence on the quantum fluctuations, i.e., temperature, fields $\to
0$, and boundaries $\to \infty$. Even then, the quantum vacuum may be thought
of as an infinity of ubiquitous virtual processes -- 
fluctuations of the quantum fields representing creations and
annihilations of wave packets (``particles'') in spacetime -- which
are compatible with Heisenberg's uncertainty principle. 

Even if the ground state realizes the naive anticipation of vanishing
field expectation values, we can probe the complex structure of the
quantum vacuum by applying external fields or boundaries etc., and
measuring the response of the vacuum to a suitable probe. For
instance, let us send a weak light beam into the box; it may interact
with the virtual fluctuations and will have finally traveled through
the box at just ``the speed of light''. If we switch on an external
magnetic field, the charged quantum fluctuations in the box are
affected and reordered by the Lorentz force. This has measurable
consequences for the speed of the light probe which now interacts with
the reordered quantum fluctuations. Thus, quantum field theory
invalidates the superposition principle of Maxwell's theory. The
quantum world creates nonlinearities and also nonlocalities
\cite{Heisenberg:1936qt,Dittrich:2000zu,Dittrich:1985yb}.

Quantum vacuum physics inspires many research branches, ranging from
mathematical physics studying field theory with boundaries and functional
determinants to applied physics where the fluctuations may eventually be used
as a building block to design dispersive forces in micro- and nanomachinery.
Many quantum vacuum phenomena such as the Casimir effect are similarly
fundamental in quantum field theory as the Lamb shift or $g-2$ experiments,
and hence deserve to be investigated and measured with the same effort. Only a
high-precision comparison between quantum vacuum theory and experiment can
reveal whether we have comprehensively understood and properly computed the
vacuum fluctuations.

In this article, I will argue that, with such a comparison, one
further step can be taken: a high-precision investigation can then
also be used to look for systematic deviations as a hint for new
physics phenomena. Similarly to $g-2$, quantum vacuum experiments can
systematically be used to explore new parameter ranges of
particle-physics models beyond the Standard Model (BSM). 

What are the scales of sensitivity which we can expect to probe?
Consider a typical Casimir experiment: micro- or mesoscopic setups
probe dispersive forces between bodies at a separation
$a=\mathcal O(\text{nm}-10\mu\text{m})$. This separation $a$ also sets
the scale for the dominant quantum fluctuation wavelengths which are
probed by the apparatus. The corresponding energy scales are of order
$\mathcal O(10\text{meV}-100\text{eV})$. As another example, consider
an optical laser propagating in a strong magnetic field of a few
Tesla. Again, the involved energy scales allow to probe quantum
fluctuations below the $\mathcal O(10\text{eV})$ scale. Therefore,
quantum vacuum experiments can probe new physics below the eV scale
and hence are complementary to accelerators. Typical candidates are
particles with masses in the meV range, i.e., {\em physics at the
  milli scale} \cite{Lindner:2007zz}. 

The particular capability of these experiments is obviously not a
sensitivity to heavy particles, but a sensitivity to light but
potentially very weakly coupled particles. In the following, I will
especially address optical experiments. Here, there are at least
two lever arms for increasing the sensitivity towards weak coupling:
Consider a laser beam entering an interaction region, say a magnetized
quantum vacuum; some photons may leave the region towards a
detector. Let us assume that the setup is such that the Standard
Model predicts zero photons in the detector; this implies that the
observation of a single photon (which is technically possible) is
already a signature for new physics. On the other hand, an incoming
beam, for instance, from an optical 1000Watt laser contains $\sim
10^{21}$ photons per second. It is this ratio of $10^{21}:1$ which
can be exploited for overcoming a weak-coupling suppression. Second,
the interaction region does not have to be microscopic as in
accelerator experiments, but can be of laboratory size (meters) or can
even increase to kilometers if, e.g., the laser light is stored in a
high-finesse cavity.

Why should we care for the milli scale at all? First of all, exploring
a new particle-physics landscape is worthwhile in itself; even if
there is no discovery, it is better to {\em know} about a
non-existence than to {\em assume} it. Second, we already know about
physics at the milli scale: neutrino mass differences and potentially
also their absolute mass is of order $\mathcal O(1-100\text{meV})$;
also, the cosmological constant can be expressed as $\Lambda\sim
(2\text{meV})^4$. A more systematic search for further particle
physics at the milli scale hence is certainly worthwhile and could
perhaps lead to a coherent picture. Third, a large number of
Standard-Model extensions not only involves but often requires -- for
reasons of consistency -- a hidden sector, i.e., a set of so far
unobserved degrees of freedom very weakly coupled to the Standard
Model. A discovery of hidden-sector properties could much more
decisively single out the relevant BSM extension than the discovery of
new heavy partners of the Standard Model. 

Optical quantum vacuum experiments can be very sensitive to new light
particles which are weakly coupled to photons. From a bottom-up
viewpoint, I will first discuss low-energy effective theories of the
Standard Model and of BSM extensions which allow for a classification
of possible phenomena and help relating optical observables with
fundamental particle properties. Subsequently, current bounds on
new-physics parameters are critically examined. In Sec.~\ref{sec3}, I
briefly describe current and future experimental setups, and discuss
recently published data. An emphasis is put on the question of how
dedicated quantum vacuum experiments can distinguish between different
particle-physics scenarios and extract information about the nature of
the involved degrees of freedom. Section \ref{sec4} gives a short
account of underlying microscopic models that would be able to
reconcile a large anomalous signal in the laboratory with
astrophysical bounds. Conclusions are given in Sec.~\ref{sec5}.

\section{Low-energy effective actions}
\label{sec2}

A plethora of ideas for BSM extensions can couple new particle
candidates to our photon. From a bottom-up viewpoint, many of these
ideas lead to similar consequences for low-energy laboratory
experiments, parameterizable by effective actions that describe the
photon coupling to the new effective degrees of freedom. In the
following, we list different effective actions that are currently
often used for data analysis. This list is not unique nor complete. 

\subsection{QED and Heisenberg-Euler effective action}

The first example is standard QED as a low-energy effective theory of
the Standard Model: if there are no light particles coupling to the
photon other than those of the Standard Model, present and near-future
laboratory experiments will only be sensitive to pure QED degrees of
freedom, photon and electron. If the variation of the involved fields
as well as the field strength is well below the electron mass scale,
the low-energy effective action is given by the lowest-order
Heisenberg-Euler effective action
\cite{Heisenberg:1936qt,Dittrich:2000zu,Dittrich:1985yb},
\begin{equation}
\fl\Gamma_{\text{HE}}=\!
     \int_x \!
      \left\{ -\frac{1}{4} F_{\mu\nu} F^{\mu\nu} + \frac{8}{45}
        \frac{\alpha^2}{m^4}\left( \frac{1}{4} F_{\mu\nu}F^{\mu\nu}
     \right)^2
     + \frac{14}{45}\frac{\alpha^2}{m^4}\left( \frac{1}{4} F_{\mu\nu}\widetilde{F}^{\mu\nu}
     \right)^2 \!+ \mathcal{O}\left(
	   {\scriptstyle \frac{F^6}{m^8}, \frac{\partial^2 F^2}{m^2}}
	   \right)\!\right\}\!,
\label{eq:HE}
\end{equation}
which arises from integrating out the ``heavy'' electron-positron
degrees of freedom to one-loop order. In addition to the Maxwell term,
the second and third term exemplify the fluctuation-induced
nonlinearities. The corresponding quantum equations of motion thus
entail a photon self-coupling. As an example, let us consider the
propagation of a laser beam with a weak amplitude in a strong magnetic
field $B$. From the linearized equations of motion for the laser beam,
we obtain a dispersion relation which can be expressed in terms of
refractive indices for the magnetized quantum vacuum
\cite{Dittrich:2000zu,Baier,Adler:1971wn}:
\begin{equation}
   { n_\|}\simeq 1+ \frac{14}{45} \frac{\alpha^2}{m^4}\,
   B^2 \sin^2 \theta_B, \quad  
   { n_\bot}\simeq 1+ \frac{8}{45} \frac{\alpha^2}{m^4}\,
   B^2 \sin^2 \theta_B, \label{eq:refind}
\end{equation}
where $\theta_B$ is the angle between the $B$ field and the
propagation direction. Most importantly, the refractive indices,
corresponding to the inverse phase velocity of the beam, depend on the
polarization direction $\|$ or $\bot$ of the laser with respect to the
$B$ field. The magnetized quantum vacuum is birefringent. As a
corresponding observable, an initially linearly polarized laser beam
which has nonzero components for both $\|$ and $\bot$ modes
picks up an {\em ellipticity} by traversing a magnetic field: the
phase relation between the polarization modes changes, but their
amplitudes remain the same. The ellipticity angle $\psi$ is given by
$\psi=\frac{\omega}{2} \ell (n_\|-n_\bot)\sin 2 \theta$, where $\theta$
is the angle between the polarization direction and the $B$ field, and
$\ell$ is the path length inside the magnetic field. 

So far, a direct verification of QED vacuum magnetic birefringence has not
been achieved; if measured it would be the first experimental
proof that the superposition principle in vacuum is ultimately violated for
macroscopic electromagnetic fields. 

Another optical observable is important in this context: imagine some
effect modifies the amplitudes of the $\|$ or $\bot$ components in a
polarization-dependent manner, but leaves the phase relations
invariant. By such an effect, a linearly polarized beam will then
effectively change its polarization direction after a passage through
a magnetic field by a {\em rotation} angle $\Delta \theta$. Since
amplitude modifications involve an imaginary part for the index of
refraction, rotation from a microscopic viewpoint is related to
particle production or annihilation. In QED below threshold $\omega<2
m$, electron-positron pair production by an incident laser is
excluded.  Only photon splitting in a magnetic field would be an
option \cite{Adler:1971wn}. However, for typical laboratory
parameters, the mean free path exceeds the size of the universe by
many orders of magnitude and hence is irrelevant.\footnote{It is amusing
to observe that it is neutrino-pair production which could be the
largest Standard-Model contribution to an optical rotation measurement
in a strong electromagnetic field \cite{Gies:2000wc}; but, of course,
it is similarly irrelevant for current and near-future experiments.}
We conclude that a sizeable signal for vacuum magnetic rotation
$\Delta \theta$ in an optical experiment would be a signature for new
fundamental physics.

\subsection{Axion-Like Particle (ALP)}

As a first BSM example, we consider a neutral scalar $\phi$ or
pseudo-scalar degree of freedom $\phi^{-}$ which is coupled to the
photon by a dimension-five operator,
\begin{equation}
\Gamma_{\text{ALP}}= \int_x \left\{-\frac{{g}}{4} \phi^{(-)} F^{\mu\nu}
\stackrel{(\sim)}{F}_{\mu\nu}
-\frac{1}{2}(\partial\phi^{(-)})^2 - \frac{1}{2}  {m_\phi}^2 \phi^{(-)}{}^2
\right\}. 
\label{eq:ALP}
\end{equation}
This effective action is parameterized by the particle's mass $m_\phi$
and the dimensionful coupling $g$. For the pseudo-scalar case, this
action is familiar from axion models \cite{Peccei:1977hh}, where the two
parameters are related, $m_\phi \sim g$. Here, we have a more general
situation with free parameters in mind which we refer to as axion-like
particles (ALP). In optical experiments in strong $B$ fields, ALPs can
induce both ellipticity and rotation \cite{Maiani:1986md}, since only
one photon polarization mode couples to the axion and the external
field: the $\|$ mode in the pseudo-scalar case, the $\bot$ mode in the
scalar case. For instance, coherent photon-axion conversion causes a
depletion of the corresponding photon mode, implying rotation. Solving
the equation of motion for the coupled photon-ALP system for the
pseudo-scalar case yields a prediction for the induced ellipticity and
rotation,
\begin{equation}
\fl
{{\Delta\theta}}^{-}\!
  = \left({\scriptstyle \frac{gB\omega}{m_\phi^2}}\right)^2
    \sin^2\!\left({\scriptstyle \frac{L m_\phi^2}{4\omega}}\right)\sin2\theta,
\,\,\,
\psi^{-}\!
=\!\! \frac{1}{2}\left({\scriptstyle
  \frac{gB\omega}{m_\phi^2}}\right)^2\left({\scriptstyle \frac{L m_\phi^2}{2\omega}}-
\sin\!\left({\scriptstyle \frac{L
    m_\phi^2}{2\omega}}\right)\right)\sin2\theta,
\label{eq:ellrotALP}
\end{equation}
for single passes of the laser through a magnetic field of length $L$.
For the scalar, we have $\Delta\theta=-\Delta\theta^{-}$,
$\psi=-\psi^{-}$. This case is a clear example of how fundamental
physics could be extracted from a quantum vacuum experiment: measuring
ellipticity and rotation signals uniquely determines the two
model parameters, ALP mass $m_\phi$ and ALP-photon coupling $g$.
Measuring the signs of $\Delta \theta$ and $\psi$ can even resolve the
parity of the involved particle.

Various microscopic particle scenarios lead to a low-energy effective action
of the type \eqref{eq:ALP}. The classic case of the axion represents an
example in which only the weak coupling to the photon is relevant and all
other potential couplings to matter are negligible. In this case, the laser
can be frequency-locked to a cavity such that both quantities are enhanced by
a factor of $N_{\text{pass}}$ accounting for the number of passes. For the
generated ALP component, the cavity is transparent. This facilitates another
interesting experimental option, namely, to shine the ALP component through a
wall which blocks all the photons. Behind the wall, a second magnetic field
can induce the reverse process and photons can be regenerated out of the ALP
beam \cite{Sikivie:1983ip}. The regeneration rate is
\begin{equation}
\label{regALPps}
\dot{N}_{\gamma\ {\rm reg}}^{(-)} = \dot{N_0}\left(
\frac{N_\text{pass}+1}{2}\right)
\frac{1}{16}\left(gBL\cos\theta\right)^4\left[
{\sin\left(\frac{L m_\phi^2}{4\omega}\right)} \Big/
{\frac{L m_\phi^2}{4\omega}}
\right]^4,
\end{equation}
where $\dot{N_0}$ is the initial photon rate, and the magnetic fields
are assumed to be identical.

In other models, such as those with a chameleon mechanism \cite{cham},
the ALP cannot penetrate into the cavity mirrors but gets reflected
back into the cavity. Whereas this has no influence on the single-pass
formulas for $\psi$ and $\Delta \theta$ in \Eqref{eq:ellrotALP}, the
use of cavities and further experimental extensions can be used to
distinguish between various microscopic models, see below.

\subsection{Minicharged Particle (MCP)}

In addition to the example of a neutral particle, optical experiments can also
search for charged particles. If their mass is at the milli scale, these
experiments can even look for very weak coupling, i.e., minicharged particles
(MCPs) \cite{Okun:1982xi}, the charge of which is smaller by a factor of
$\epsilon$ in comparison with the electron charge. If the MCP is, for
instance, a Dirac spinor $\psi_\epsilon$, the corresponding action is
\begin{equation}
\Gamma_{\text{MCP}}= - \bar\psi (i \fss{\partial} + \epsilon e
\fsl{A}) \psi + m_\epsilon \bar\psi \psi
\label{lintDsp},
\end{equation}
where we again encounter two parameters, $\epsilon$ and the MCP mass
$m_{\epsilon}$. At a first glance, the system looks very similar to
QED. However, since the particle mass $m_\epsilon$ can be at the milli
scale or even lighter, the weak-field expansion of the
Heisenberg-Euler effective action for slowly varying fields
\eqref{eq:HE} is no longer justified. Both field strength as well as
laser frequency can exceed the electron mass scale with various
consequences \cite{Gies:2006ca}: the laser frequency can be above the
pair-production threshold $\omega>2m_\epsilon$ such that a rotation
signal becomes possible. Second, there is no perturbative ordering
anymore as far as the coupling to the $B$ field is concerned, hence
the MCP fluctuations have to be treated to all orders with respect to
$B$. All relevant information is encoded in the polarization tensor
corresponding to an MCP loop with two photon legs and an
infinite number of couplings to the $B$ field which is well known from
the QED literature
\cite{Dittrich:2000zu,Dittrich:1985yb,Toll:1952rq}. Explicit results
are available in certain asymptotic limits, for instance, for the
rotation,
\begin{equation}
\fl  \Delta \theta \simeq \frac{1}{12} \frac{\pi}{\Gamma(\frac{1}{6})
  \Gamma(\frac{13}{6})} \left(\!\frac{2}{3}\!\right)^{\frac{2}{3}}
  {\epsilon}^2 \alpha ({m_\epsilon} \ell) 
  \left( \frac{{m_\epsilon}}{\omega}\right)^{\frac{1}{3}} 
  \left( \frac{{\epsilon} e{B}}{{m_\epsilon}^2}
  \right)^{\frac{2}{3}}, \quad\text{for}\,\, \frac{3}{2}
  \frac{\omega}{m_{\epsilon}} \frac{\epsilon e B}{m_\epsilon^2}\gg 1, 
\label{eq:rotMCP}
\end{equation}
which is valid above threshold and for a high number of
allowed MCP  Landau levels. Similar formulas exist for ellipticity or the
case of spin-0 MCPs \cite{Ahlers:2006iz}. Note that this rotation
becomes independent of $m_\epsilon$ in the small-mass limit, such that
\Eqref{eq:rotMCP} apparently implies a sensitivity to arbitrarily
small masses. In practice, this sensitivity is limited for other
reasons: for instance, once the associated Compton wavelength $\sim
1/m_\epsilon$ becomes larger than the size of the magnetic field, the
constant-field assumption, which is often used for calculating the
polarization tensor, is no longer valid. The rotation depends on the
size of the magnetic field, the scale of which acts as a cutoff for
the sensitivity towards smaller masses; e.g. for $\ell\simeq 1$m, the
MCP mass should satisfy $m_\epsilon\gg 0.2\mu$eV. Let me stress that
the computation of polarization tensors in inhomogeneous fields is a
challenge for standard methods and remains an interesting question for
future research. Progress may come from modern worldline techniques
\cite{Gies:2001zp,Gies:2005bz}. 

\subsection{Paraphotons}

In addition to neutral scalars or weakly charged particles, we may
also consider additional (hidden) gauge fields which interact weakly
with the photon. A special coupling is provided by gauge-kinetic
mixing which occurs only between abelian gauge fields, hence involving
a second photon, i.e., a paraphoton $\gamma'$ \cite{Okun:1982xi},
\begin{equation}
\Gamma_{\gamma\gamma'}= -\frac{1}{4} F_{\mu\nu}F^{\mu\nu} 
-\frac{1}{4}{F'_{\mu\nu} F'{}^{\mu\nu} }
-\frac{1}{2} {\chi}\, F^{\mu\nu} {F_{\mu\nu}'}-
\frac{1}{2} {\mu}^2 A_\mu' A'{}^\mu,
\label{eq:para}
\end{equation}
with a mixing parameter $\chi$ and a paraphoton mass term $\mu$.
Without the mass term, the kinetic terms could be diagonalized by a
non-unitary shift $A'_\mu \to \hat{A}'_\mu - \chi A_\mu$ which would
decouple the fields at the expense of an unobservable charge
renormalization. The mass term does not remain diagonal by this shift,
such that observable $\gamma\gamma'$ oscillations arise from mass
mixing in this basis. The pure paraphoton theory is special in the
sense that $\gamma\gamma'$ conversion is possible without an external
field and is not sensitive to polarizations. For instance, the
conversion rate after a distance $L$ is given by
$\mathcal{P}_{\gamma\to\gamma'}= 4 \chi^2 \sin^2 \frac{\mu^2 L}{4
\omega}$. Therefore, paraphotons can be searched for in future
light-shining-through-walls experiments \cite{Ahlers:2007rd}. Below, we
discuss microscopic scenarios in which paraphotons and MCPs naturally
occur simultaneously.

\subsection{Bounds on low-energy effective parameters}

Many different observations seem to constrain the parameters in the
effective theories listed above. The strongest constraints typically
come from astrophysical observations usually in combination with
energy-loss arguments. Consider, for instance, the ALP low-energy
effective action \eqref{eq:ALP}. Assuming that it holds for various
scales of momentum transfer, we may apply it to solar physics. Thermal
fluctuations of electromagnetic fields in the solar plasma, giving
rise to non-vanishing $F^{\mu\nu} \stackrel{(\sim)}{F}_{\mu\nu}$, act
as a source for $\phi^{(-)}$ ALPs. In absence of other sizeable
interactions, ALPs escape the solar interior immediately and
contribute to stellar cooling. A similar argument for the
helium-burning life-time of HB stars leads to a limit $g\lesssim
10^{-10}$GeV${}^{-1}$ for ALP masses in the eV range and below
\cite{Raffelt:2006cw}. Monitoring actively a potential axion flux from
the sun as done by the CAST experiment even leads to a slightly better
constraint for ALP masses $<0.02$eV \cite{Andriamonje:2007ew}.

Astrophysical energy-loss arguments constrain also MCPs
\cite{Davidson:2000hf}: for instance, significant constraints on
$\epsilon$ come from helium ignition and helium-burning lifetime of HB
stars, resulting in $\epsilon\leq 2\times 10^{-14}$ for $m_{\epsilon}$
below a few keV.

Without going into detail, let us stress that all these bounds on the
effective-action parameters depend on the implicit assumption that the
effective actions hold equally well at solar scales as well as in the
laboratory. But whereas solar processes typically involve momentum
transfers on the keV scale, laboratory quantum vacuum experiments
operate with much lower momentum transfers, a typical scale being
$\mu$eV. In other words, the above bounds can only be applied to
laboratory experiments, if one accepts an extrapolation of the
underlying model over nine orders of magnitude. In fact, it has been
shown quantitatively how the above-mentioned bounds have to be
relaxed, once a possible dependence of these effective-action
parameters, e.g., on momentum transfer, temperature, density or other
ambient-medium parameters, is taken into account
\cite{Jaeckel:2006xm}. This observation indeed provides for another
strong imperative to perform well-controlled laboratory experiments.

Previous laboratory experiments have also produced more direct
constraints on the effective action parameters. For instance, the best
laboratory bounds on MCPs previously came from limits on the branching
fraction of ortho-positronium decay or the Lamb shift
\cite{Davidson:2000hf,Gluck:2007ia}, resulting in $\epsilon \lesssim
10^{-4}$. Similarly, pure laboratory bounds on ALP parameters used to
be much weaker than those from astrophysical arguments.

\section{From optical experiments to fundamental particle properties}
\label{sec3}

A variety of quantum vacuum experiments is devoted to a study of
optical properties of modified quantum vacua. The BFRT experiment
\cite{Cameron:1993mr} has pioneered this field by providing upper
bounds on vacuum-magnetically induced ellipticity, rotation as well as
photon regeneration. Improved bounds for ellipticity and rotation have
recently been published by the PVLAS collaboration
\cite{Zavattini:2007ee}.\footnote{The new data is no longer compatible
with the PVLAS rotation signal reported earlier
\cite{Zavattini:2005tm}. Nevertheless, this artifact deserves the
merit of having triggered the physics-wise well justified rapid
evolution of the field which we are currently witnessing.} Further
polarization experiments such as Q\&A \cite{Chen:2006cd} and BMV
\cite{Robilliard:2007bq} have also already taken and published data.

The PVLAS experiment uses an optical laser ($\lambda=1064$nm and
532nm) which is locked to a high-finesse Fabry-Perot cavity
($N=\mathcal O(10^5)$) and traverses an $L= 1$m long magnetic field of up
to $B=5.5$Tesla. Owing to the high finesse, the optical path length
$\ell$ inside the magnet effectively increases up to several 10km. 

The improved PVLAS bounds for ellipticity and rotation can directly be
translated into bounds on the refractive-index and
absorption-coefficient differences, $\Delta n=n_\| - n_\bot$,
\begin{equation}
\fl |\Delta n(B=2.3 \text{T})|\leq 1.1 \times 10^{-19}
    /\text{pass}, \quad  
|\Delta \kappa({B}=5.5 \text{T})| \leq  5.4 \times
    10^{-15} \text{cm}^{-1} .\label{eq:PVLASres}
\end{equation}
As an illustration, an absorption coefficient of this order of
magnitude would correspond to a photon mean free path in the magnetic
field of the order of a hundred times the distance from earth to sun,
demonstrating the quality of these laboratory experiments. 

These bounds imply new constraints, e.g., for the ALP parameters,
$g\simeq 4\times 10^{-7}$GeV${}^{-1}$ for $m_\phi<1$meV. More
importantly, for MCPs, we find $\epsilon\lesssim 3\times 10^{-7}$ for
$m_\epsilon < 30$meV. This bound is indeed of a similar size as a
cosmological MCP bound which has recently been derived from a
conservative estimate of the distortion of the energy spectrum of the
cosmic microwave background \cite{Melchiorri:2007sq}. Hence,
laboratory experiments begin to enter the parameter regime which has
previously been accessible only to cosmological and astrophysical
considerations. 

Imagine an anomalously large signal, say, for ellipticity $\psi$ and
rotation $\Delta\theta$ is observed by such a polarization experiment,
thereby providing evidence for vacuum-magnetic birefringence and
dichroism. How could we extract information about the nature of the
underlying particle-physics degree of freedom? The two data points for
$\psi$ and $\Delta\theta$ can be translated into parameter pairs $g$
and $m_\phi$ for ALPs, $\epsilon$ and $m_\epsilon$ for MCPs, etc.,
leaving open many possibilities. As already mentioned above, a
characteristic feature is the sign of $\psi$ and $\Delta\theta$; i.e.,
identifying the polarization modes $\|$ or $\bot$ as fast or slow
modes reveals information about the microscopic properties
\cite{Ahlers:2006iz}: e.g., a pseudo-scalar ALP goes along with
$\Delta\kappa,\Delta n>0$, whereas a scalar ALP requires
$\Delta\kappa,\Delta n<0$. A mixed combination, say $\Delta\kappa>0,
\Delta n<0$, would completely rule out an ALP, leaving a spinor MCP as
an option, etc.

Another test would be provided by varying the experimental parameters
such as length or strength of the magnetic field, or the laser
frequency \cite{Ahlers:2006iz}. For instance, an ALP-induced rotation
exhibits a simple $B^2$ dependence, the nonperturbative nature of
MCP-induced rotation results in a $B^{2/3}$ law,
cf. Eqs.~\eqref{eq:ellrotALP}, \eqref{eq:rotMCP}.

The underlying degree of freedom can more directly be identified by
special-purpose experiments that probe a specific property of particle
candidate. The light-shining-through-walls experiment is an example for such a
setup. A magnetically-induced photon regeneration signal in such an
experiment would clearly point to a weakly interacting ALP degree of freedom;
 the outgoing photon polarization would distinguish between scalar
($\bot$ mode) or pseudo-scalar ($\|$ mode) ALPs. For this reason, a
number of light-shining-through-walls experiments is currently being built or
already taking data: ALPS at DESY \cite{Ehret:2007cm}, LIPSS at JLab
\cite{Afanasev:2006cv}, OSQAR at CERN \cite{OSQAR}, and GammeV at Fermilab
\cite{GammeV}. PVLAS will shortly be upgraded accordingly, and BMV has already
published first results \cite{Robilliard:2007bq}, yielding a new bound
$g\lesssim1.3\times10^{-6}$GeV${}^{-1}$ for $m_\phi\lesssim 2$meV.

MCPs do not contribute to a photon regeneration signal, since pair-produced
MCPs inside the magnet are unlikely to recombine behind the wall and produce a
photon.\footnote{Photon regeneration can still be a decisive signal for
  models with both MCPs and paraphotons \cite{Ahlers:2007rd}.} A
special-purpose quantum vacuum experiment for MCP production and detection has
been suggested in \cite{Gies:2006hv}: a strong electric field, e.g., inside an
RF cavity can produce an MCP dark current by means of the nonperturbative
Schwinger mechanism \cite{Heisenberg:1936qt,Gies:2005bz}. A first signature
could be provided by an anomalous fall-off of the cavity quality factor (the
achievable high-quality factor of TESLA cavities already implies the bound
$\epsilon\lesssim 10^{-6}$ \cite{Gies:2006hv}). Owing to the weak interaction,
the MCP current can pass through a wall where a dark current detector can
actively look for a signal.

In the case of a strongly interacting ALP, photon regeneration behind a wall
would not happen either, since the wall would block both photons and generated
ALPs. A special example is given by chameleon models which have been developed
in the context of cosmological scalar fields and the fifth-force problem
\cite{cham}. Somewhat simplified, chameleons can be viewed as ALPs with a
varying mass that increases with the ambient matter density. As a result,
low-energy chameleons which are initially produced in vacuo by photon
conversion in a magnetic field cannot penetrate the end caps of a vacuum
chamber and are reflected back into the chamber. After an initial laser pulse,
the chameleons can be re-converted into photons again inside the magnetized
vacuum; this would result in an afterglow phenomenon which is characteristic
for a chameleonic ALP \cite{Gies:2007su}. First estimates indicate that the
chameleon parameter range accessible to available laboratory technology is
comparable to scales familiar from astrophysical stellar energy loss
arguments, i.e., up to $g\sim 10^{10}$GeV for $m_\phi\lesssim 1$meV. Afterglow
measurements are already planned at ALPS \cite{Ehret:2007cm} and GammeV
\cite{GammeV}.

In the near future, also quantum vacuum experiments could be realized
that involve strong fields generated by a high-intensity laser; for a
concrete proposal aiming at vacuum birefringence, see
\cite{Heinzl:2006xc} and also \cite{DiPiazza:2006pr}. As major
differences, laser-driven setups can generate field strengths that
exceed conventional laboratory fields by several orders of magnitude.
The price to be paid is that the spatial extent of these high fields
is limited to a few microns. We expect that laser-driven experiments
can significantly contribute to MCP searches in the intermediate-mass
range whereas ALP and paraphoton searches, which are based on a
coherence phenomenon, typically require a spatially sizeable field.

Both spatially extended as well as strong fields are indeed available in
the vicinity of certain compact astrophysical objects. Also cosmic
magnetic fields though weak may be useful due to their extreme spatial
extent. For suggestions how to exploit these fields as a probe for
fundamental physics, see, e.g.,
\cite{Dupays:2005xs,Mirizzi:2007hr,DeAngelis:2007yu}.

\section{Microscopic models}
\label{sec4}

So far, we argued that quantum vacuum experiments do not only
serve as a probe for fundamental physics and BSM extensions, but also are
required to provide for model-independent information about potential
weakly coupled light degrees of freedom. Nevertheless, in the case of
a positive anomalous experimental signal a puzzle of how to reconcile
this signal with astrophysical bounds would persist on the basis of the 
low-energy effective actions discussed above. A resolution of this
puzzle has to come from the underlying microscopic theory that
interconnects solar scales with laboratory scales. 

A number of ideas has come up to separate solar physics from
laboratory physics; for a selection of examples, see
\cite{Masso:2006gc,Mohapatra:2006pv,Jain:2006ki,Foot:2007cq,cham,%
Antoniadis:2007sp}. A general feature of many ideas is to suppress the
coupling between photons and the new particle candidates at solar
scales by a parameter of the solar environment such as temperature,
energy or momentum transfer, or ambient matter density. A somewhat
delicate alternative is provided by new particle candidates that are
strongly interacting in the solar interior, resulting in a small mean
free path (similar or smaller than that of the photons!), such that
they do not contribute to the solar energy flux \cite{Jain:2006ki}.

A paradigmatic example for a parametrical coupling suppression is
given by the Masso-Redondo model \cite{Masso:2006gc} which, in
addition to resolving the above puzzle, finds a natural embedding in
string-theory models \cite{Abel:2006qt}. As a prerequisite, let us
consider the paraphoton model of \Eqref{eq:para} and include a
hidden-sector parafermion $h$ which couples only to the paraphoton
$A'$ with charge $e_{\text{h}}$ and interaction $e_{\text{h}} \bar h
\fsl{A}' h$. After the shift $A'_\mu \to \hat{A}'_\mu - \chi A_\mu$
which diagonalizes the kinetic terms, the parafermion acquires a
coupling to our photon: $-\chi e_{\text{h}} \bar h \fsl{A}
h$. Since $\chi$ is expected to be small, we may identify $-\chi
e_{\text{h}}=\epsilon e$. As a result, the hidden-sector fermion
appears as minicharged with respect to our photon. The bottom line is
that a hidden sector with further U(1) fields and correspondingly
charged particles automatically appear as MCPs for our photon if
these further U(1)'s mix weakly with our U(1). However, if the
paraphoton is massive the coupling of on-shell photons to
parafermions is suppressed by this mass $\mu$, since the on-shell
condition cannot be met by the massive paraphoton. 

The Masso-Redondo model now involves two paraphotons, one massless and
one massive, with opposite charge assignments for the
parafermions. The latter charge assignment indeed cancels the
parafermion-to-photon coupling at high virtuality (as, e.g., for the
photon plasma modes in the solar interior), implying that solar
physics remains unaffected. At low virtualities such as in the
laboratory, the massive paraphoton decouples which removes the
cancellations between the two U(1)'s. A photon-paraphoton system is
left over in which the parafermions indeed appear as MCPs with respect
to electromagnetism. In this manner, the astrophysical bounds remain
satisfied, but laboratory experiments could discover unexpectedly
large anomalous signatures. 

In fact, hidden sectors also involving further U(1)'s and
correspondingly charged matter as required for the Masso-Redondo
mechanism cannot only be embedded naturally in more fundamental
models, but are often unavoidable in model building for reasons of
consistency.

\section{Conclusions}
\label{sec5}
 
Quantum vacuum experiments such as those involving strong external
fields can indeed probe fundamental physics. In particular, optical
experiments can reach a high precision and thereby constitute an ideal
tool for searching for the hidden sector of BSM extensions containing
weakly-interacting and potentially light degrees of freedom at the
milli scale. A great deal of current experimental activity will soon
provide for a substantial amount of new data which will complement
particle-physics information obtained from accelerators. 

From a theoretical viewpoint, many open problems require a better
understanding of fluctuations of light degrees of freedom, the small
mass of which often inhibits conventional perturbative ordering
schemes. Modern quantum-field theory techniques for external-field
problems such as the worldline approach
\cite{Schubert:2001he,Gies:2001zp,Gies:2005bz} will have to be used
and developed further hand in hand with experimental progress in
probing the quantum vacuum.

I would like to thank M.~Ahlers, W.~Dittrich, D.F.~Mota, J.~Jaeckel,
J.~Redondo, A.~Ringwald, D.J.~Shaw for collaboration on the topics
presented here. It is a pleasure to thank M.~Bordag and his team for
organizing the QFEXT07 workshop and creating such a stimulating
atmosphere. This work was supported by the DFG under contract No.
Gi 328/1-4 (Emmy-Noether program).

\end{document}